\documentclass[prb]{revtex4}

\usepackage{graphicx}
\usepackage{dcolumn}
\usepackage{bm}

\begin{document}


\title{On Possibilities of Studying of Supernova Neutrinos at BAKSAN}

\author{G.V. Domogatsky}
\affiliation{%
Institute for Nuclear Research RAS, Moscow\\
}%

\author{V.I. Kopeikin}%
\author{L.A. Mikaelyan}%
\author{V.V. Sinev}%
\affiliation{
Russian Research Center ``Kurchatov Institute"\\
}%

\date{\today}

\begin{abstract}
We consider the possibilities of studying a supernova collapse neutrino burst at Baksan Neutrino Observatory (Institute for Nuclear Research, 
Russian Academy of Sciences) using the prposed 5-kt target-mass liquid scintillation spectrometer. 
Attention is given to the influence of mixing angle ${\theta}_{13}$ on the expected rates and spectra of neutrino events.

\end{abstract}

\maketitle

\section*{Introduction}

Neutrino fluxes of natural origin are known to carry information both about their sources 
inaccessible to direct observation and about the  properties of the neutrino themselves.

We consider a program of research on natural low-energy ($<$ 100 MeV) antineutrino fluxes 
using a large scintillation spectrometer to be installed at Baksan
Neutrino Observatory (BNO) of the Institute for Nuclear Research, 
Russian Academy of Sciences, at a depth of $\sim$5000 m.w.e. 
The planned spectrometer target mass is about 5 kt.

The main research directions and goals are the following:

(1) Neutrino geophysics.
Determining the radiogenic component of the Earth's heat flow through the detection of antineutrinos 
emitted by daughter uranium and thorium decay products; detrmining the total abundance of these elements 
in the Earth. 

Testing the hypothesis that a chain fission reaction is burning at the Earth's centre (``georeactor'').

(2) Neutrino Astrophysics.
Studying the Supernova (SN) expolsion dynamics by recording the neutrino burst intensity and energy spectrum. 
Searching for an isotropic flux of antineutrinos accumulated in the Universe over several billion years 
during SN explosions and the formation of neutron stars and black holes.

The first section of the research program has been schematically considered in recent publications [1-4].

This paper is devoted to the possibilities of studying the neutrino burst when cores of high-mass stars 
gravitationally collapse  to produce SNe.

The burst of ``Thermal'' neutrinos begins simultaneously with gravitational collapse of the iron core
of a high-mass ($M\ge 8M_{\odot}$) star 
and lasts for about 20 s. All six types of active neutrinos are produced in a hot core:
$\nu_{e}, \bar{\nu_{e}},\nu_{\mu}, \bar{\nu_{\mu}},\nu_{\tau}$ and $\bar{\nu_{\tau}}$. 
These neutrinos carry away the vast bulk of the gravitational energy, $\sim 3\times 10^{53}$ erg , released during the collapse. 
Only several tens of minutes or several hours later does the exposion reach the surface, and a Supernova seen with the naked eye 
and by the methods of optical, X-ray, radio and gamma-ray astronomy flares up in the sky. 
On competition of the explosion, the bulk of the stellar matter is dispersed in space and a neutron star (or a black hole) 
is left at the place of the core.

By neutrino burst onset, the inner part of the core becomes opaque to neutrinos because of its
high density. Before their escape, the neutrinos are multiply scattered, absorbed and reemitted. The surface from which 
the neutrinos can leave the core is called a neutrinosphere. Because of the difference in interaction cross sections,
the neutrinosphere radii differ for defferent types neutrinos. 
The $\nu_{\mu}, \bar{\nu_{\mu}},\nu_{\tau}$  and  $\bar{\nu_{\tau}}$ neutrinospheres (radius $\sim$30 km) are deepest, 
the $\bar{\nu_{e}}$ neutrinosphere ($\sim$50 km) follows next, and the latter is followed by the $\nu_{e}$ neutrinosphere
($\sim$70 km). This explains the predicted difference in mean energies of the neutrinos escaping from the core. 
The larger the radius, the lower the mean energy. However, as calculations show, all types of neutrinos divide the 
collapse energy carried away by them approximately evenly.

A short pulse ($\sim10^{-2}$ s) of electron neutrinos, $\nu_{e}$, produced via neutronization of the core that reached the stability
(Chandrasekhar) limit is emitted immediately before the ``thermal" pulse. These $\nu_{e}$'s with
energies 15-20 MeV, carry away 5-10\% of the energy released during the collapse.

Studying the neutrino bursts in labratories is a powerful tool for investigating the energetics and dynamics of collapsing stars
and the properties of neutrinos themselves.

The existence of a neutrino burst that accompanies collapse was first pointed out in 1965 [5]. 
Intensive calculations of the SN explosion dynamics and neutrino emission were performed in succeeding years [6-9]. 
There are a number of detailed reviews (see [10, 11] and references therein). A neutrino burst detection method 
was also suggested in 1965 [12]: a statistically significant series of signals must be observed
in a massive low-background neutrino detector. However a SN explosion is a very rare event (one SN explosion in $\sim$30 years
is expected in our Galaxy). Therefore, the authors of [12] suggested synchronizing the operation of various 
detectors to increase reliability and information content. In our days there exists The International SuperNova Early Warning System
(SNEWS) that combines all of the detectors capable of detecting SN neutrinos is currently operating. 

The neutrinos produced during collapse play an important role in nucleosynthesis [15-17]. The observed abundances of several 
light elements ($^{9}$Be, $^{11}$B, $^{19}$F, and others), and the abundances of so-called bypassed elementes heavier than iron, 
can be explained in terms of neutrino reactions within $\sim$1000 km of the stellar core centre.

The signal from SN 1987A in the Large Magellanic Cloud was recordered on February 23, 1987, with three neutrino 
detectors - KII in Japan, IMB in the USA and the Baksan detector. This signal was analyzed in details
in a number of papers (see, e.g. [18, 19]). A total of 24 neutrinos were recordered in the above detectors (the 
explosion occured at 
a distance of 50 kpc, which a factor 5 larger than the mean distance from the Earth to the stars of our Galaxy). 
On the whole, the picture obtained is consistent with the outlined neutrino burst scenario, although poor statistics severely 
limits the information content of the analysis.

Note that SN neutrino burst detection is an oscillation experiment with a baseline of $\sim$30 thousand 
light years and that the neutrinos produced in the star traverse regions that are a factor of $\sim 10^9$ denser 
than the matter the solar center. The flavour composition of neutrinos incident on the detector depends on 
the mixing parameter $\sin^2{\theta}_{13}$ and on the type of mass hierarchy (normal or inversed) 
[20, 21]. The conversions $\bar{\nu_{\mu}},\bar{\nu_{\tau}}\Longrightarrow \bar{\nu_{e}}$ 
open the possibility of studying the intensities and energies of 
$\bar{\nu_{\mu}}$ and $\bar{\nu_{\tau}}$, whose signal is virtually impossible to isolate directly by currently available methods.

Below, we (1) schematically describe the neutrino detector, (2) give the characteristics of a ``standard" thermal 
neutrino burst, (3) calculate the spectra and rates of expected neutrino events in the detector, (4) consider the effects 
of oscillations on the number and spectrum of neutrino events and, (5), in conclusion, discuss our results.

\section{Neutrino detector}

The planned detector will consist of three concentric zones. The central spherical zone abouit
 23 m in diameter is filled with a liquid organic scintillator and serves as a target for
neutrinos. The target is separated from the second zone filled with a nonscintillating oil 
by a strong transparent film. Photomultiplier tubes (PMTs) are installed on the surface 
of the second zone and view the target through $\sim$4-meter layer of liquid. The outer zone
of the detector is an anticoincidence zone. It is separated from 
second zone by an opaque metall structure filled with water (or oil) and is viewed by PMTs
that record Cerenkov radiation from cosmic muons and showers. The outer size of the detector 
is $\sim$33 m.

The PMT photocathodes viewing the target cover about 25\% of the surface. The expected
signal is 80-100 photoelectrons per 1 MeV. 

The main difference between the detector under consideration and the already existing
LVD [21] and SNO [22] detectors, 
which are capable of detecting a neutrino burst, is a larger target mass, a good energy 
resolution, a low detection threshold, and the ability to detect ``terrestrial" 
antineutrinos. The detector under construction is closest in structure to the KamLAND 
array [23], differing from it by a larger target mass and a deeper position underground. 
The recently suggested LENA project [24] has a program similar to ours, but it 
supposes constructing a detector with much larger target mass.

The main (in time) mode of operation of the spectrometer is a set of statistics of events
generated by terrestrial antineutrinos produced via $^{238}$U and $^{232}$Th decays and 
by the hypothetical georeactor [1-4]. The number of useful events of this type is  
$\sim$500 per year. In the mode of neutrino burst detection, the expected number 
of events can reach several thousand over a period of $\sim$20 s.

\begin{figure}
\includegraphics{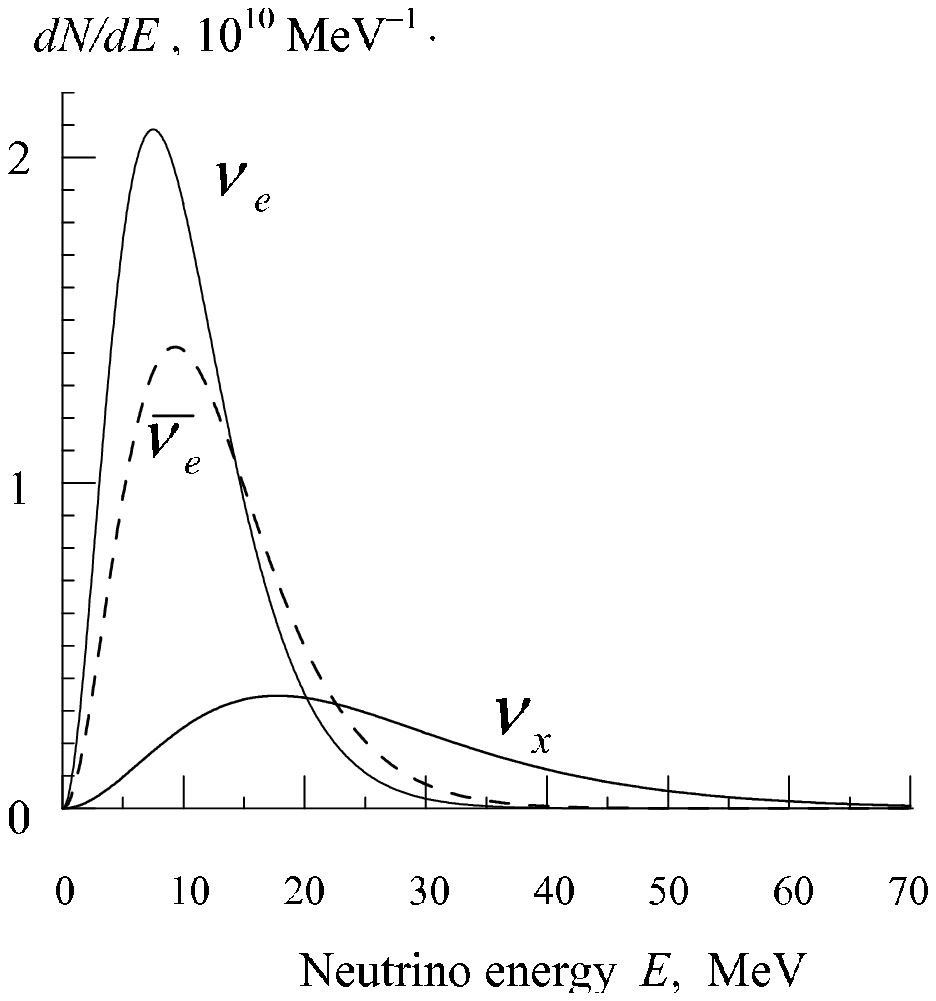}
\caption{\label{fig:epsart} Spectra of the neutrinos incident on the detector (without oscillations),
 $\nu_x$ = $\nu_{\mu}, \bar{\nu_{\mu}},\nu_{\tau}, \bar{\nu_{\tau}}$.}
\end{figure}

\section{The thermal neutrino burst}

We calculate the expected effects in the detector based on the quantitative characteristics 
of the neutrino burst given below.

(1) Thermal neutrinos carry away an energy of $3\times10^{53}$ erg, which is divided 
evenly between the six flavours:
$\nu_{e},\bar{\nu_{e}},\nu_{\mu},\bar{\nu_{\mu}},\nu_{\tau}$ and $\bar{\nu_{\tau}}$.

(2) The neutrino energy distribution is described by Fermi-Dirac distribution 
with an additional factor that provides a faster falloff of the spectrum at high
neutrino energies:

\begin{equation}
N(E)dE = \frac{C\cdot E^2dE}{1+e^{E/T}}e^{-{\alpha}(E/T)^2},
\end{equation} 
where $C$ is the normalization constant, $E$ and $T$ $-$ are the energy and temperature 
(in MeV) and $\alpha << 1$ is a dimensionless parameter.

The numerical values of the quantities that characterize spectrum (1) were taken from [7]
(see Table I).

The last two lines in Table I give the mean neutrino energies of neutrino calculated
from distribution (1) and the burst-time-integrated flux 
densities of various types of neutrinos on the detector. 
The distance from the collapsing star to the detector was assumed to be 10 kpc (remind 
that 1 pc $\approx 3.086\times10^{18}$ cm).

The spectra of the neutrinos incident on the detector are shown in figure 1. 
Note a peculiarity of our model that is important for subsequent analysis: 
the $\nu_{x}$ neutrinos $=(\nu_{\mu},\bar{\nu_{\mu}},\nu_{\tau},\bar{\nu_{\tau}})$ 
have an appreciably harder 
spectrum than $\nu_{e}$ and $\bar{\nu_{e}}$.

\begin{table}[t]
\caption{Characteristics of the thermal neutrino burst (the SN distance is 10 kpc, without oscillations)}
\label{table:1}
\vspace{10pt}
\begin{tabular}{l|c|c|c}
\hline
Neutrino flavour  & $\nu_{e}$ & $\bar{\nu_{e}}$ & $\nu_{x}^*$ \\
\hline
Carried-away energy $W$, $10^{52}$ erg & 5 & 5 & $5^{**}$ \\
\hline
Temperature $T$, MeV  & 3.5 & 4.5 & 8 \\
\hline
Dimensionless parameter $\alpha$  & 0.01 & 0.02 & 0$^{**}$ \\
\hline
Mean neutrino energy $<E>$/MeV  & 10.3 & 12.5 & 25.2$^{**}$ \\
\hline
Flux $\nu$ on detector  $N_{0i}$, $10^{11}$ /cm$^2$  & 2.537 & 2.088 & 1.035$^{**}$ \\
\hline
\end{tabular}\\[2pt]
{\small $\nu_{x}=\nu_{\mu},\bar{\nu_{\mu}},\nu_{\tau},\bar{\nu_{\tau}}$}\\
{\small relates to each $\nu_{x}$ neutrino}
\end{table} 

\section{Neutrino reactions in the detector}

In this section we consider the effects expected in the detector without neutrino oscillations. (see the second line in Table II). 
We assume that the useful target volume contains $4\times10^{32}$ hydrogen atoms, 
$2\times10^{32}$ $^{12}$C atoms, and $16\times10^{32}$ electrons.

\begin{table}[t]
\caption{Effects of oscillations on the numbers of neutrino events (normal mass hierarchy)}
\label{table:1}
\vspace{10pt}
\begin{tabular}{l|c|c|c}
\hline
Reaction  & without & LMA MSW & LMA MSW \\
      & oscillation & $\sin^{2}{\theta}_{13}>10^{-3}$ & $\sin^{2}{\theta}_{13}<10^{-5}$ \\
\hline
$\bar{\nu_{e}}+p \rightarrow n + e^{+}$ & 1157 & 1479 & 1479  \\
\hline
$\bar{\nu_{e}} \ + \ ^{12}{\rm C}\rightarrow \ ^{12}{\rm B} + e^{+}$ & 14.4 & 35.5 & 35.5  \\
\hline
$\nu_{e} \ + \ ^{12}{\rm C}\rightarrow \ ^{12}{\rm N} + e^{-}$ & 5.8 & 132 & 93.5  \\
\hline
$\sum$ CC on $^{12}$C & 20.2 & 167 & 129  \\
\hline
$\sum (\nu_{i},^{12}$C) $\rightarrow ^{12}$C+15.1$\gamma$ & 236 & 236 & 236  \\
\hline
$\sum (\nu_{i},e) \rightarrow (\nu_{i},e)$ & 70.6 & 62.2 & 61.4  \\
\hline
\end{tabular}\\[2pt]
\end{table} 

3.1. The inverse beta-decay reaction in which a free proton converts into a neutron and positron, 

\begin{equation}
\bar{\nu_{e}}+p \rightarrow n + e^{+}.
\end{equation}
has the largest cross section. 
The $\bar{\nu_{e}}$ threshold is 1.806 MeV. Light signals correlated in time and space are recorded in the detector from the positron and  
from the $\gamma$-ray photons of the capture of the neutron from reaction (2) by scintillator hydrogen. 
The mean neutron lifetime before the capture is $\sim$200 $\mu$s. 
The energy released by the positron in the scintillator $E_{vis}$ (MeV), is related to the energy of the incoming 
$\bar{\nu_{e}}$ by:
$$
E_{vis}=E_{\nu}-R_n-0.78, \eqno(2a)
$$
where $R_n$ is the neutron recoil energy. The energy averaged over the positron escape angles, $R_n$, increases quadratically 
with incoming neutrino energy and reaches $\sim$2 MeV at $E_{\nu}$= 50 MeV. 
When calculating the cross sections and spectra, we used results from [25]. A total of about 1200 events of the reaction 
under consideration is expected over the burst.

3.2. The reactions of electron antineutrino and neutrino interaction with carbon in the channels of charged currents,

\begin{equation}
\bar{\nu_{e}} \ + \ ^{12}{\rm C}\rightarrow \ ^{12}{\rm B} + e^{+} (E_{th}=13.9 \ {\rm MeV});
\end{equation}
$$
\quad^{12}{\rm B} \rightarrow ^{12}{\rm C}+e^{-}(T_{1/2}=20.2 ms),
$$
$$
\nu_{e} \ + \ ^{12}{\rm C}\rightarrow \ ^{12}{\rm N} + e^{-} (E_{th}=17.8 \ {\rm MeV}); \eqno (3a) 
$$
$$
\qquad ^{12}{\rm N} \rightarrow ^{12}{\rm C}+e^{+}(T_{1/2}=11 ms),
$$
produce $^{12}$B and ($^{12}$N), which decay with the emission of an electron and positron, respectively. Spatially correlated 
delayed coincidences and energies escaping charged leptons are recorded in the detector.  
Because of high reaction thresholds, the number of events is very small (Table II). 
Ther cross sections for reactions (3 and 3a) were found in [26] and are presented in Figure 2. 

\begin{figure}
\includegraphics{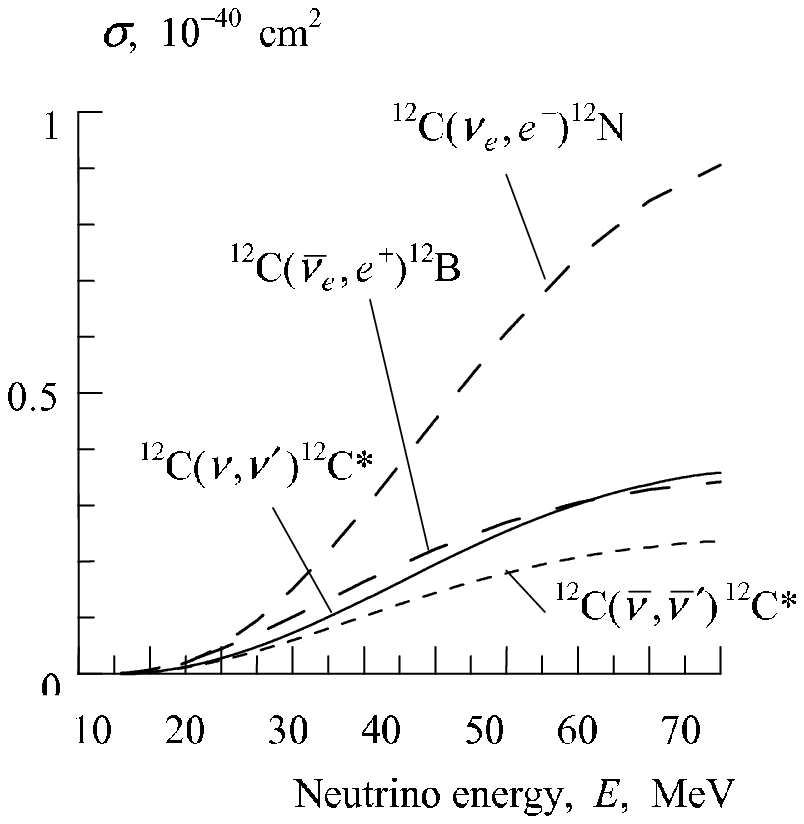}
\caption{\label{fig:epsart} Cross sections for interactions of neutrinos with $^{12}$C in the channels of neutral and charge 
currents.}
\end{figure}

Neutrinos also can be scattered inelastically by $^{12}$C,
$$
\nu_{e} \ + \ ^{12}{\rm C}\rightarrow \ ^{12}{\rm C} + \gamma, \quad E_{\gamma}=15.1 \ {\rm MeV}, \eqno  (3b)
$$
to produce a monochromatic line of isolated photons recorded by the detector.
Because of low energies, the total contribution from $\nu_{e}$ and $\bar{\nu_{e}}$ to their total number does not exceed 5\% and 
the hard $\nu_{x}$ neutrinos play a dominant role.

3.3. About 70 isolated recoil electrons are expected from the $(\nu,e)$ scattering reaction. 
The cross sections for these processes are presented in figure 3. 

\begin{figure}
\includegraphics{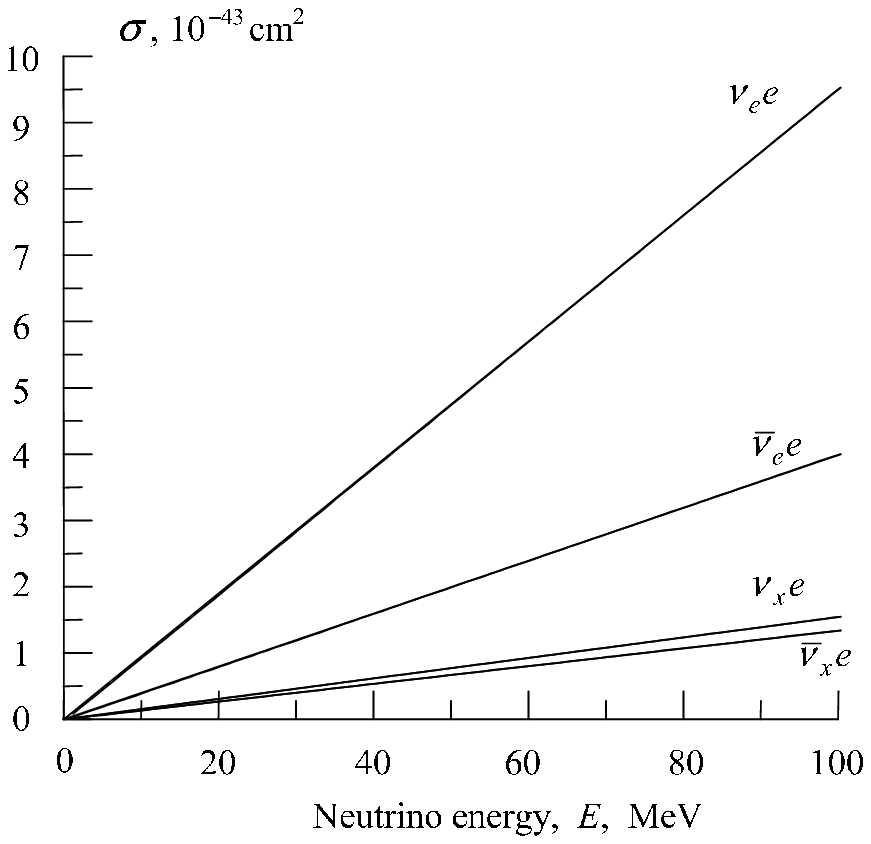}
\caption{\label{fig:epsart} Cross sections of neutrino interaction on electron.}
\end{figure}

\section{Neutrino after oscillations}

4.1. The oscillations of the neutrinos produced in a stellar core on their way to the detector were considered in detail in [20].

Because of the $\bar{\nu_{e}}\Longleftrightarrow \bar{\nu_{\mu}},\bar{\nu_{\tau}}$, and $\nu_{e}\Longleftrightarrow \nu_{\mu},\nu_{\tau}$ 
transitions the intensities and spectra of the neutrinos reaching the detector change noticeably compared to the initial ones given in 
table I, which were used to obtain the results of the previous section in this paper. The probabilities of these oscillation transitions 
depend on several factors.

(1) On the amplitudes $U_{e1}, U_{e2}$ and $U_{e3}$, with which mass states $m_{1}, m_{2}, m_{3}$ enter into the 
superposition of $\bar{\nu_{e}}$ and $\nu_{e}$ flavour states. In recent years, it has been established (see e.g. 
[27, 28]), that large mixing angle (LMA) case occurs in Nature. 
In the subsequent we assume that

\begin{equation}
{\rm tan}^2{\theta}_{12}\equiv \mid U_{e2}/U_{e1}\mid ^2=0.40, 
\end{equation}
$$
\mid U_{e1}\mid ^2=0.7, \ \mid U_{e2}\mid ^2=0.3
$$
The small, but important quantity $|U_{e3}|^2 \equiv  \sin^2{\theta}_{13}$ remains unceratin. The following experimental constraint
was obtained for it: 
$$
|U_{e3}|^2 \equiv  \sin^2{\theta}_{13} \le 0.03 \ {\rm (CHOOZ \ [29])} \eqno (4a)
$$

(2) On the influence of stellar matter influence and on the presence and nature of resonances (adiabatic, non-adiabatic) in the stellar matter 
with a variable density (the Mikheev-Smirnov-Wolfenstein, (MSW) effect).

(3) On the type of mass hierarchy (normal, $m_{3}^2 >> m_{1}^2$, or inverted $m_{3}^2 << m_{1}^2$). 
For normal and inverted mass hierarchies, respectevely, $\nu_{e}$ and $\bar{\nu_{e}}$ undergo the resonant conversions.

Below, we will consider normal mass hierarchy, making one exclusion that clearly demonstrates the role  
of type of mass hierarchy . For brevity, we will disregard the influence of the Earth's matter on 
the neutrino way to the detector.

4.2. Let us now consider the changes in the observed picture of events to which the oscillations 
$\bar{\nu_{e}}\Longleftrightarrow \bar{\nu_{\mu}},\bar{\nu_{\tau}}$ lead.

The spectrum of $\bar{\nu_{e}}$ neutrinos incident on the detector, $d^{osc}N_{\bar{\nu_{e}}}/dE_{\bar{\nu_{e}}}$, is given relation
\begin{eqnarray}
\frac{d^{osc}N_{\bar{\nu_{e}}}}{dE_{\bar{\nu_{e}}}}= |U_{e1}|^2\frac{dN_{\bar{\nu_{e}}}}{dE_{\bar{\nu_{e}}}}+
|U_{e2}|^2\frac{dN_{\nu_x}}{dE_{\nu_x}}+|U_{e3}|^2\frac{dN_{\nu_x}}{dE_{\nu_x}}, 
\end{eqnarray}
i.e.,
\begin{eqnarray}
\frac{d^{osc}N_{\bar{\nu_{e}}}}{dE_{\bar{\nu_{e}}}}= |U_{e1}|^2\frac{dN_{\bar{\nu_{e}}}}{dE_{\bar{\nu_{e}}}}+
(1-|U_{e1}|^2)\frac{dN_{\nu_x}}{dE_{\nu_x}} \approx 0.7\frac{dN_{\bar{\nu_{e}}}}{dE_{\bar{\nu_{e}}}}+
0.3\frac{dN_{\nu_x}}{dE_{\nu_x}} \nonumber
\end{eqnarray}
Relation (5) implies that 70\% of the initial soft $\bar{\nu_{e}}$'s reach the detector in an unchanged form and 30\% of 
the initial hard $\nu_{x}$ are added to them in the form of $\bar{\nu_{e}}$'s. 
As a result, the number of events of inverse beta-decay reaction (2) increases (see Table II) and a
hard component appears in the spectrum of positrons from this reaction (figure 4a). In addition the number of events of the reaction
$\bar{\nu_{e}} \ + \ ^{12}{\rm C}\rightarrow \ ^{12}{\rm B} + e^{+}$ increases by several times. Of course the total number of
neutrinos is conserved and 30\% of the initial $\bar{\nu_{e}}$ turn into $\bar{\nu_{x}}$.

4.3. In the process $\nu_{e}\Longleftrightarrow \nu_{\mu},\nu_{\tau}$, a resonance whose nature
depends on the small parameter $|U_{e3}|^2 = \sin^2{\theta}_{13}$ is observed in stellar matter.

In $\sin^{2}{\theta}_{13}>10^{-3}$, then only a negligible fraction of the initial soft 
$\nu_{e}$ reaches the detector, in fact, 
the entire flux of hard $\nu_{e}$ is determined by the $\nu_{x}\rightarrow \nu_{e}$ transition.
\begin{equation}
\frac{d^{osc}N_{\nu_{e}}}{dE_{\nu_{e}}}= |U_{e3}|^2\frac{dN_{\nu_{e}}}{dE_{\nu_{e}}}+
(1-|U_{e3}|^2)\frac{dN_{\nu_x}}{dE_{\nu_x}} \approx \frac{dN_{\nu_x}}{dE_{\nu_x}}
\end{equation}
As a result, the number of events of the reaction 
$\nu_{e} \ + \ ^{12}{\rm C}\rightarrow \ ^{12}{\rm N} + e^{-}$
increases by a factor of $\sim$ 20 (Table II). The total number of recoil electrons from 
neutrino scattering by electrons changes insignificantly.

If, alternatively, $\sin^{2}{\theta}_{13}<10^{-5}$, then 30\% , (i.e., $|U_{e2}|^2$) 
of the initial soft $\nu_{e}$ reaches the detector, and 70\% ($1-|U_{e2}|^2$) of 
the initial hard $\nu_{x}$ turn into $\nu_{e}$. 
Thus, for $\sin^{2}{\theta}_{13}>10^{-3}$, the $\nu_{e}$ spectrum contains more hard neutrinos 
than for $\sin^{2}{\theta}_{13}<10^{-5}$.

\begin{figure}
\includegraphics{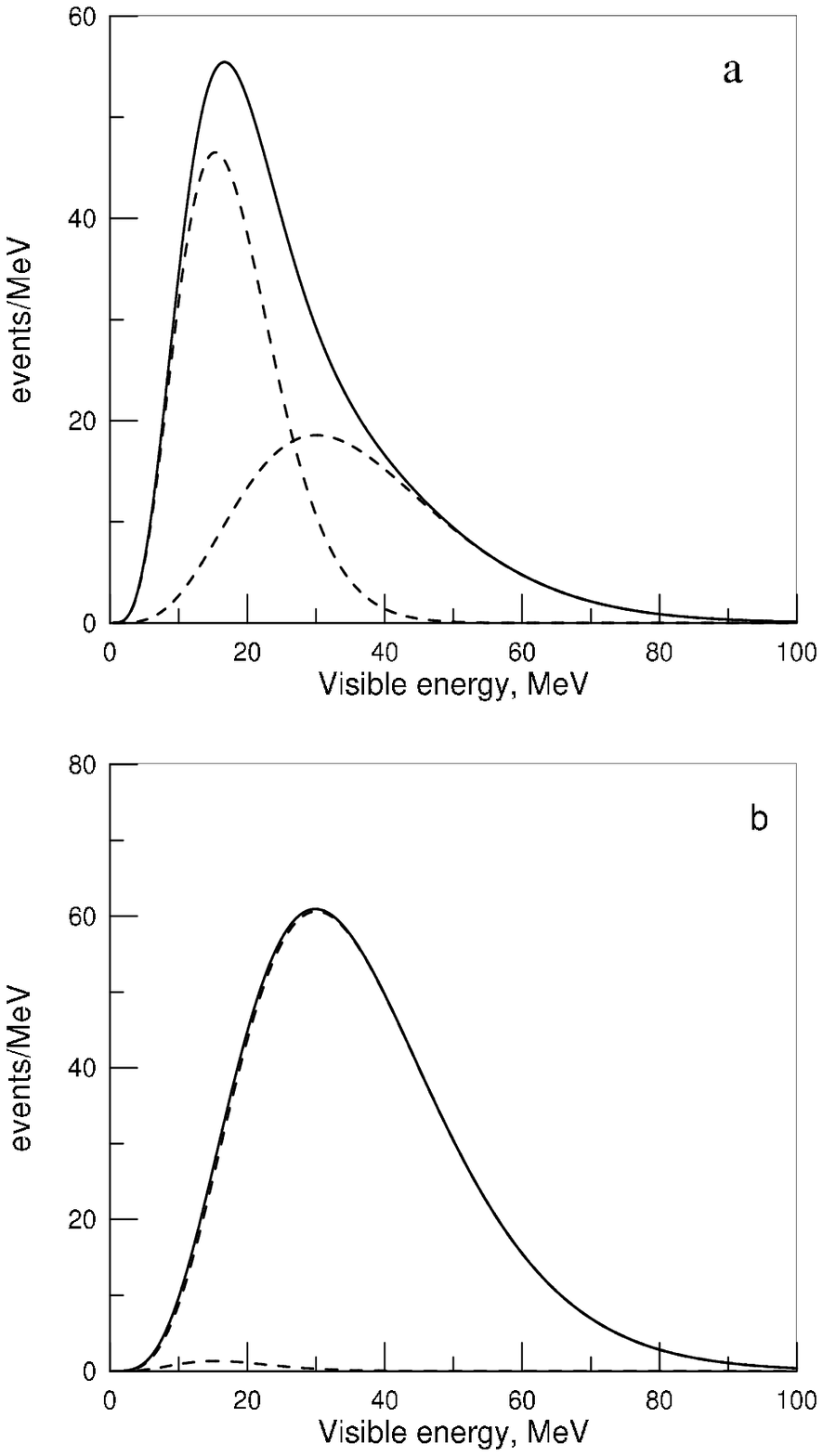}
\caption{\label{fig:epsart} Spectra of the positrons from the inverse beta-decay reaction (2) 
with oscillations:
(a) Spectrum for normal mass hierarchy, (b) spectrum for inverted mass hierarchy, 
under the condition $\sin^{2}{\theta}_{13}>10^{-3}$.}
\end{figure}

4.4. Let us also consider the case of inverted neutrino mass hierarchy,
$m_{3}^2 << m_{1}^2$. 
The MSW resonance now occurs during 
$\bar{\nu_{e}}\Longleftrightarrow \bar{\nu_{\mu}},\bar{\nu_{\tau}}$ transitions 
and depends on $\sin^{2}{\theta}_{13}$. 

If $\sin^{2}{\theta}_{13}>10^{-3}$, then the incident flux 
$d^{osc}N_{\bar{\nu_{e}}}/dE_{\bar{\nu_{e}}}$ 
is determined almost entirely by the hard $\bar{\nu_{x}}$that turned into $\bar{\nu_{e}}$:
\begin{equation}
\frac{d^{osc}N_{\bar{\nu_{e}}}}{dE_{\bar{\nu_{e}}}}\approx \frac{dN_{\nu_{x}}}{dE_{\nu_{x}}}.
\end{equation}
As a result, the total number of events of reaction (2) reaches $\sim$2200 
and an intense peak of hard $e^+$ will be observed in the positron spectrum 
(see figure 4b).

If, alternatively, $\sin^{2}{\theta}_{13}<10^{-5}$, then the spectrum of $\bar{\nu_{e}}$ 
incident on the detector is identical to spectrum (5) that has already been
considered above and the positron spectrum does not differ from the shown at figure 4a.

\begin{table}[t]
\caption{Characteristics of the thermal burst neutrino spectrum form various sources}
\label{table:1}
\vspace{10pt}
\begin{tabular}{l|c|c|c}
\hline
$<E_{\nu_{e}}>$, MeV & $<E_{\nu_{x}}>$, MeV & cite & year  \\
\hline
12.5 & 25.2 & [7] & 1980  \\
15.4 & 21.6 & [9] & 1998  \\
11.4 & 14.1 & [32] & 2003  \\
15.4 & 15.7 & [33] & 2003  \\
$\approx 10-12^*$ & $\approx 25$ & [34] & 2005  \\
\hline
\end{tabular}\\[2pt]
{\small The mean value for $\bar{\nu_{e}}$ and $\nu_{e}$}
\end{table} 

\section{Conclusion}

The parameter $|U_{e3}|^2 = \sin^2{\theta}_{13}$ plays an important role in the physics of
weak interactions. CP-invariance violations in the neutrino sector are possible only 
if $|U_{e3}|^2 \neq 0$. 
Ways of improving the experiment 
limitations on the parameter and achieving a better sensitivity were suggested immediately
after constraint (4b)  was obtained on the CHOOZ reactor [30]. 
At present, preparations are being made for new experiments in the field of the reactors
of many countries (see [31] for a brief review). 
These experiments should be able improve result (4b) by a factor of 5-7. 
The dependence of the spectrum and number of neutrino events on $\sin^2{\theta}_{13}$ 
(Table II, figure 4) gives hope that this parameter will be determined with a good sensitivity
in future astrophysical observations.

Detection of a neutrino burst can give more information than mentioned above. 

The processes that take place in the core of a star within the first seconds after its 
collapse are very complex. 
It is not surprising that specialists in this field have failed to reach a consensus
of opinion on the spectral composition of the emitted neutrinos even after
several decades of intensive calculations. Above we used a model ([7], figure 1), in which
$\nu_{x}=\nu_{\mu},\bar{\nu_{\mu}},\nu_{\tau},\bar{\nu_{\tau}}$ have appreciably higher mean 
energies than $\nu_{e}$ and $\bar{\nu_{e}}$. At the ssame time, there are studies suggesting 
that the $\nu_{x}$ and $\bar{\nu_{e}}$ energies are close [32] or almost equal 
[33], while [34] gives data similar to those in [7] (see table III). 

What is the dynamics of the core interior temperature within the first seconds after the core
collapse? 
How does the neutrino burst flare up and fade out? How the neutrino gas cools down in the burst 
time? 
And, finally, does a rapidly rotating star actually collapses in two stages separated in time,
as explained in detail in [35]?
Only analysis of a future neutrino burst can give answers to these questions.

\section*{Acknowledgments }

We thank O.G. Ryazhskaya, D.K. Nadyozhin and A.Yu. Smirnov for the discussions and 
valuable advices and A. Strumia for the discussion on calculation the nuances of the
inverse beta-decay reaction cross section.
The work was supported by Russian Foundation for Basic Research (project 06-02-16024a).


\begin{thebibliography}{14}
\bibitem{domog1} G.V. Domogatsky, V.I. Kopeikin, L.A. Mikaelyan, V.V. Sinev, Phys.At.Nucl., \textbf{68}, 69 (2005);
ArXiv: hep-ph/0401221.
\bibitem{domog3} G.V. Domogatsky, V.I. Kopeikin, L.A. Mikaelyan, V.V. Sinev, Phys.At.Nucl., \textbf{68}, 234 (2005);
ArXiv: hep-ph/0403155.
\bibitem{domog2} G.V. Domogatsky, V.I. Kopeikin, L.A. Mikaelyan, V.V. Sinev, Phys.At.Nucl., \textbf{69}, 43 (2006);
ArXiv: hep-ph/0409069.
\bibitem{domog4} G.V. Domogatsky, V.I. Kopeikin, L.A. Mikaelyan, V.V. Sinev, Phys.At.Nucl., \textbf{69}, 1894 (2006); ArXiv: hep-ph/0411163.
\bibitem{zeldov} Ya. B. Zeldovich, O.Kh. Gusseinov, Reports of Academy of Science USSR (rus), \textbf{162}, 791, (1965).
\bibitem{arnet} Arnett W.D., Can. J. Phys., \textbf{44}, 2553 (1966).
\bibitem{nadiuj} D.K. Nadyozhin, N.V. Otroschenko, Sov.Astron., \textbf{24}, 47 (1980).
\bibitem{bower} Bowers R., Wilson J. R., Astrophys. J., \textbf{263}, 366 (1982).
\bibitem{totani} Totani T. et al., Astrophys. J., \textbf{496}, 216 (1998).
\bibitem{cei} F. Cei, Int. J. Mod. Phys., \textbf{A17}, 1765 (2002).
\bibitem{bethe} H.A. Bethe, Rev. Mod. Phys., \textbf{62}, 801 (1990).
\bibitem{domog5} G. Domogatsky, G. Zatsepin, Proc. 9 Int. Cosmic Rays Conf.,  London, Vol. 2, 1030 (1965).
\bibitem{antoni} P. Antonioli et al., New J. Phys., \textbf{6}, 114 (2004); astro-ph/0406214.
\bibitem{scholb} K. Scholberg, Nucl. Phys. B (Proc. Suppl.), \textbf{91}, 331 (2001). 
\bibitem{domog6} G. Domogatsky, D.K. Nadyozhin, Mont. Not. R.Astron.Soc., \textbf{178}, 33 (1977).
\bibitem{domog7} G. Domogatsky, R.A. Eramzhyan, D.K. Nadyozhin, Astroph. Space Sci., \textbf{58}, 273 (1978).
\bibitem{woosly} S.E. Woosley et al., Astrophys. J., \textbf{365}, 272 (1990).
\bibitem{dadyk} V.L. Dadykin, G.T. Zatsepin, O.G. Ryazhskaya, Sov.Phys.Usp., \textbf{32}, 459 (1989).
\bibitem{imshen} V. Imshennik, D. Nadyozhin, Sov. Sci. Rev. E 8, part 1, 156 (1989).
\bibitem{dighe} A.S. Dighe, A.Yu. Smirnov, Phys. Rev., \textbf{D62}, 033007 (2000); hep-ph/9907423.
\bibitem{agliet} M. Aglietta et al., Nucl. Phys. Proc. Suppl., \textbf{110}, 410 (2002); astro-ph/0112312.
\bibitem{virtue} C. J. Virtue (for the SNO Collab.), Nucl. Phys. B (Proc. Suppl.), \textbf{100}, 326 (2001).
\bibitem{kaml} The KamLAND Collab., Phys. Rev. Lett., \textbf{94}, 081801 (2005).
\bibitem{katrin}  K.A. Hochmuth, F. v. Feilitzsch et al., hep-ph/0509136.
\bibitem{strum} A. Strumia, F. Vissani, Phys. Lett., \textbf{B564}, 42 (2003).
\bibitem{fukigi} M. Fukugita, Y. Kohyama, K. Kubodera, Phys. Lett., \textbf{B212}, 139 (1988).
\bibitem{smirnov} A.Yu. Smirnov, hep-ph/0402264.
\bibitem{fogli} G. Fogli, E. Lisi et al., hep-ph/0506083.
\bibitem{fukigi} M. Apollonio et al., Phys. Lett., \textbf{B466}, 415 (1999).
\bibitem{mikael2} L. Mikaelyan, Nucl. Phys. B (Proc. Suppl.), \textbf{87}, 284 (2000); 
\textbf{91}, 120 (2001).
\bibitem{goodm} M. Goodman, hep-ph/0501206.
\bibitem{thomps} Thompson T., Borrows A., Pinto P., Astrophys. J., \textbf{592}, 434 (2003).
\bibitem{keil} Keil M., Raffelt G., Janka H., Astrophys. J., \textbf{590}, 971 (2003).
\bibitem{nadyoj2} D. Nadyozhin, V. Imshennik, astro-ph/0501002.
\bibitem{imshenn} V. Imshennik, O. Ryazhskaya, Astron. Lett., \textbf{30}, 14 (2004); astro-ph/0401613.

\end{thebibliography}
\end{document}